\documentstyle[12pt,epsf]{article}
\newcommand {\beq}{\begin{equation}}
\newcommand {\eeq}{\end{equation}}
\newcommand {\beqa}{\begin{eqnarray}}
\newcommand {\eeqa}{\end{eqnarray}}
\newcommand {\beqan}{\begin{eqnarray*}}
\newcommand {\eeqan}{\end{eqnarray*}}
\newcommand {\n}{\nonumber \\}

\newcommand {\eq}[1]{eq.~(\ref{#1})}

\newcommand {\del}{\partial}

\begin{document}

\begin{titlepage}
\renewcommand{\thefootnote}{\fnsymbol{footnote}}

\begin{flushright}
\begin{tabular}{l}
KEK-TH-654 \\
NSF-ITP-99-120\\
\end{tabular}
\end{flushright}

\vspace{2cm}
\begin{center}
         {\Large Scaling Behaviors of Branched Polymers} \\
\end{center}
\vspace{1cm}

\begin{center}

           Hajime A{\sc oki}$^{1)}$\footnote
           {
e-mail address : haoki@itp.ucsb.edu, JSPS abroad research fellow},
           Satoshi I{\sc so}$^{2)}$\footnote
           {
e-mail address : satoshi.iso@kek.jp},
           Hikaru K{\sc awai}$^{3)}$\footnote
           {
e-mail address : hkawai@gauge.scphys.kyoto-u.ac.jp}
          and Yoshihisa K{\sc itazawa}$^{2)}$\footnote
           {
e-mail address : kitazawa@post.kek.jp}\\
\vspace{1cm}
        $^{1)}$ {\it Institute for Theoretical Physics, UCSB}\\
               {\it Santa Barbara, , CA 93106, USA} \\
        $^{2)}$ {\it High Energy Accelerator Research Organization (KEK),}\\
               {\it Tsukuba, Ibaraki 305, Japan} \\
        $^{3)}$ {\it Department of Physics, Kyoto University,} \\
                 {\it Kyoto 606-8502, Japan}\\
\end{center}

\vfill

\begin{abstract}
\noindent
We study the thermodynamic behavior of branched polymers.
We first study random walks in order
to clarify the thermodynamic relation between the canonical ensemble
and the grand canonical ensemble.
We then show that correlation functions for branched polymers are given by
those for $\phi^3$ theory  with a single mass insertion,
not those for the $\phi^3$ theory themselves.
In particular, the two-point function behaves as $1/p^4$, not as $1/p^2$,
in the scaling region.
This behavior is consistent with the fact that the Hausdorff dimension
of the branched polymer is four.
\end{abstract}
\vfill
\end{titlepage}
\vfil\eject

\section{Introduction}
\setcounter{equation}{0}
\setcounter{footnote}{0}
Branched polymers are the simplest generalization of the
random walk and have been studied extensively\cite{ambj}\cite{frohlich}.
It is of great importance not only in statistical physics but
also in particle physics, in particular, for understanding
the critical behavior of random surfaces and quantum
gravity\cite{yoneya}\cite{kato}\cite{periwal}.
Our recent interest in branched
polymers arose in our  attempt to formulate
superstring theory nonperturbatively.
In the paper \cite{AIKKT}, we have studied the dynamics of the type
IIB  matrix model in such a framework
(\cite{IKKT} and see \cite{review} for review).
In our matrix model approach, the eigenvalues of matrices
are interpreted as space-time coordinates.
In these investigations, we find the system of branched polymers
in a simple approximation.
Although it is far from the flat four dimensional manifold, branched
polymers share the same (fractal) dimensionality four with our space-time.
It might be the first indication that superstring can explain the
dimensionality
of our space-time.
\par
In this  paper, we comment on a field theoretic description
of branched polymers.
It is well-known
that a system of random walks is described by a free scalar field
theory if there is no  effect of self-avoidance.
Similarly it is widely believed that the system of branched polymers
is described by a scalar field theory with a three-point coupling,
that is, $\phi^3$ scalar field theory.
We will, however,  show that it is not so by treating the
universal part of the partition function carefully.
The system of branched polymers without self-avoidance can be
exactly solvable by introducing the grand canonical ensemble
and using the so called Schwinger-Dyson technique.
In order to extract the correct large $N$ limit ($N$ is the system size)
or the thermodynamic limit, we have to check
that the grand canonical ensemble is dominated
by the larger size system.
In other words, we have to extract the universal part.

Our claim is that
we need a single
mass insertion in each $m$-point correlation function
of the $\phi^3$ scalar field theory
in order to describe $m$-point correlation functions in branched polymers.
Mass insertion here means a change of a propagator
in each $m$-point function from
an ordinary one, $1/p^2$, to $1/p^4$.
In particular, the two-point function behaves as $1/p^4$, not $1/p^2$.
Let us count the number of points which lie within distance $R$ from a fixed
point
in $d(>4)$ dimensions.
In the random walk, it can be estimated as $R^2$ by using the two-point
function.
We obtain $R^4$ for branched polymers by using the $1/p^4$ type propagator.
So our finding are consistent with the claim that branched polymers are four
dimensional fractals.
A multi-point correlation function is given by a sum of graphs of the
corresponding correlation function
for $\phi^3$ scalar field theory at tree level with
a single mass insertion in each graph.
\par
Our main results were announced in \cite{BPR}.
In this paper we would like to give a fuller account of our results
by providing more detailed derivations and explanations.
The organization of this paper is the following.
In section 2, we review random walks
in order to clarify the relation between the canonical and
grand canonical ensemble.
In section 3, we investigate branched polymers.
First we define a canonical ensemble for a system of branched polymers
(sec. 3.1) and then introduce grand canonical ensembles (sec. 3.2).
We emphasize that the definition of a grand canonical ensemble is not
unique.
In sec. 3.3, we solve Schwinger-Dyson equations for the conventional
grand canonical ensemble  and obtain the results
which correspond to the correlation functions of a scalar $\phi^3$ theory.
In sec. 3.4, we consider the thermodynamic limit of the correlation
functions
and obtain the correct universal behavior of them.
In section 4, we give a physical interpretation why the propagator
behaves $1/p^4$ in branched polymers.
Section 5 is devoted to conclusions and discussions.
We have two appendices A and B which derive the partition function and the
two
point function in the canonical ensemble of branched polymers.

\section{Random Walks}
\setcounter{equation}{0}
In this section, we give a brief introduction to random walks
in order to clarify the thermodynamic relation between
the canonical and grand canonical ensemble.
\par
The canonical partition function with the system size $N$ is  given by
\beq
Z_N= \int \prod_{i=1}^{N} d^d y^i
\prod_{i=1}^{N-1} f(y^i-y^{i+1})= V (\hat{f}(0))^{N-1}
\eeq
where $V$ is the total volume of the system,
and $f(x)$ is a function assigned to each bond and
damps sufficiently fast at long distance compared to the typical
length scale $a_{0}$.
$\hat{f}(p)$ is its Fourier transform:
\beq
\hat{f}(p) = \int d^d x\; e^{ipx} f(x).
\eeq
For example, we can take $f(x)= exp(-(x/a_0)^2/2)$.
\par
Correlation functions for density
$\rho(x) = \sum_{i=1}^N \delta^{(d)}(x-y^i)$
can be easily calculated.
One point function becomes
\beq
<\rho(x)>_N = {1 \over Z_N} \int \prod_{i=1}^{N} d^d y^i
\prod_{i=1}^{N-1} f(y^i-y^{i+1}) \sum_{i=1}^N \delta^{(d)}(x-y^i)
= {N \over V}.
\eeq
Two-point function is defined as
\beq
<\rho(x^1) \rho(x^2) >_N = {1 \over Z_N} \int \prod_{i=1}^{N} d^d y^i
\prod_{i=1}^{N-1} f(y^i-y^{i+1}) \sum_{i=1}^N \delta^{(d)}(x^1-y^i)
\sum_{j=1}^N \delta^{(d)}(x^2-y^j)
\eeq
and its Fourier transformation is given by
\beqa
\hat{g}^{(2)}_{N}(p)
&=& \int d^d x <\rho(x) \rho(0) >_N e^{-ipx} \n
&=&  {1 \over Z_N}
\left(
          \sum_{i , j} (\hat{f}(p))^{|i-j|} (\hat{f}(0))^{N-|i-j|-1}
           \right) \n
 &=& {1 \over V} \sum_{i , j}
 \left(   {\hat{f}(p) \over \hat{f}(0)} \right)^{|i-j|}  \n
&=& {2 \over V} \sum_{s=0}^{N-1}
 (h(p))^s (N-s) - {N \over V} \n
&=& {2 N\over V (1-h(p))}
\left( 1-{h(p)(1-h(p)^N) \over N (1-h(p))}  \right) - {N \over V} \n
&=& \frac{2N}{VH(p)}\left(1-\frac{1-e^{-NH(p)}}{NH(p)} \right)
\left(1+O\left(\frac{1}{N},H(p)\right)\right),
\eeqa
where
\beqa
h(p)&\equiv&\hat{f}(p)/\hat{f}(0)=exp(-(a_0 p)^2/2),\\
H(p)&\equiv& 1-h(p) = (a_0 p)^2/2 + \cdot \cdot \cdot.
\eeqa
We can approximate the two-point function as
\beq
\hat{g}^{(2)}_{N}(p)
= {2N \over V }{1 \over H(p)}(1-\frac{1}{NH(p)}+ O({1 \over (NH(p))^2}))
\label{eq:rw2pfc}
\eeq
in the following scaling region:
\beq
{1 \over N} <  H(p) \ll 1,
\eeq
or
\beq
N^{-1/2}/a_0 < p \ll 1/a_0.
\eeq
The scaling region is between the ultraviolet cutoff scale $a_0$
and the infrared scale of the system extent,
which is given by $\xi=a_0 N^{1/2}$.
\par
The two-point correlation function is also calculated in the
grand canonical ensemble.
A conventional  grand canonical partition function is given as
\beq
Z_{\kappa_0}=
 \sum_{N=1}^{\infty} \kappa_0^{N} Z_N = {V \kappa_c \kappa_0
\over \kappa_c -\kappa_0}
\label{RW:partition0}
\eeq
where $\kappa_0$ is fugacity and $\kappa_c = (\hat{f}(0))^{-1}$.
For later convenience, we first define an
unnormalized two-point function by
\beqa
\hat{G}_{\kappa_0}^{(2)}(p) &=& \sum_{N=1}^{\infty} \kappa_0^{N}
\sum_{i, j} (\hat{f}(p))^{|i-j|} (\hat{f}(0))^{N-|i-j|-1} \n
&=&
{ \kappa_0 \over  (1-\kappa_0  )^2 }
{2 \over (1-\kappa_0 \hat{f}(p))}
-\frac{\kappa_0}{(1-\kappa_0 \hat{f}(0))^2}\n
&=& { \kappa_0 \kappa_c^3 \over  (\kappa_c-\kappa_0)^2 }
{2 \over (\kappa_c -\kappa_0 h(p))}
-\frac{\kappa_0 \kappa_c^2}{(\kappa_c-\kappa_0)^2}.
\label{RW:correlation0}
\eeqa
Hence the normalized two-point function becomes
\beqa
\hat{g}^{(2)}_{\kappa_0}(p) & \equiv &
{\hat{G}_{\kappa_0}^{(2)}(p) \over  Z_{\kappa_0}} \n
&=& {1 \over V (1-\kappa_0 \hat{f}(0))}{2 \over (1-\kappa_0 \hat{f}(p))}
-\frac{1}{V} \frac{\kappa_c}{\kappa_c-\kappa_0}\n
&=& {2 \kappa_c \over V \delta \kappa}
{1 \over (1-h(p)) + {\delta \kappa \over \kappa_c} h(p)}
-\frac{1}{V} \frac{\kappa_c}{\delta \kappa}\n
& = &   {2 \kappa_c \over V \delta \kappa}
{1 \over H(p) + {\delta \kappa \over \kappa_c}}
\left(1+O\left(\frac{\delta\kappa}{\kappa_c}, H(p) \right) \right)
\eeqa
where
$\delta \kappa = \kappa_c - \kappa_0$.
Since $\langle N \rangle = \kappa_c  / \delta \kappa$,
the correlation function behaves as that of  massive scalar particles
\beq
\hat{g}^{(2)}(p)_{\kappa_0}
\sim {2 \langle N \rangle \over V} {1 \over H(p) +
1/ \langle N \rangle }
\label{RW:final}
\eeq
and agrees with the result in the canonical ensemble calculation.
This result  gives the correlation length
$\xi=a_0 N^{1/2}$, which indicates
 the Hausdorff dimension of random walk $d_{H}=2$.
\par
Here we give two different definitions of Hausdorff dimensions.
The first one is defined in terms of
the relation between the system size $N$ and
the extent of the system.
The infrared behavior of the above two-point function
shows that the correlation damps rapidly over the length scale
$\xi=a_0 N^{1/2} $. Since the extent of the system
$L \sim \xi$ is proportional to
$N^{1/2}$, the Hausdorff dimension is
given by $d_H^{(1)}=2 $
where $d_H^{(1)}$ is defines as
\beq
L = a_0 N^{1/d_H^{(1)}}.
\label{Hdim1}
\eeq
\par
The second definition
is to use the behavior of the correlation function at much shorter
length scale than the system size $L=\xi$.
In $d$-dimensional coordinate space,
the density correlation behaves as
\beq
g^{(2)}(x) \sim {exp(-m |x|) \over |x|^{d-2} }
\eeq
where $m = 1/\xi$. If $|x| \ll 1/m $, the mass term can be
neglected and $g^{(2)}(x) \sim {1 \over |x|^{d-2}}$.
The total number of points within a ball of radius $R$ ($R \ll L$)
around a certain point is evaluated as
\beq
N(R) = \int^{R} d^{d}x g^{(2)}(x) \sim \left( {R \over a_0} \right)^2
\eeq
and  gives the Hausdorff dimension $d_H^{(2)} =2$.
Here the definition of $d_H^{(2)}$ is
\beq
N(R) =  \left( {R \over a_0} \right)^{d_H^{(2)}}.
\label{Hdim2}
\eeq
The second definition of the Hausdorff dimension is determined
only through  the behavior of correlation functions in the
scaling region $a_0 \ll x \ll \xi$ and has nothing to do with
the behaviors near the infrared cutoff.
Hence, it is a more appropriate definition than the first one from
the thermodynamic viewpoint.
Of course, it is quite natural that we should obtain the same
dimension $d_H$ from the both definitions
since the extent of the system is approximately
evaluated at the length $L$ where $N(L)=N$.
\par
Finally in this section, we comment on generalized types of
grand canonical ensembles.
Since the motivation of introducing grand canonical ensembles
is to reproduce the same thermodynamic quantities as those
in the canonical one, we may assign different weights in summing
over different $N$.
We define generalized grand canonical ensembles by
\beqa
Z_{\kappa_0,l}&=& \sum_{N=1}^{\infty}N^l \; \kappa_0^N \; Z_N, \n
\hat{G}^{(2)}_{\kappa_0,l}(p)
&=& \sum_{N=1}^{\infty}N^l \; \kappa_0^N \; \hat{G}^{(2)}_N(p),
\eeqa
where $\kappa_0$ is  fugacity.
The criterion for a `good' grand canonical ensemble is such that
we can take the correct thermodynamic limit, or, in other words,
we can correctly take the  universal part in the sum over $N$.
That is, the correlation functions in the grand canonical ensemble at the
critical value of fugacity should reproduce
those in the canonical ensemble for large $N$:
\beq
\lim_{N \to \infty } g_{N}^{(2)}
= \lim_{\kappa_0 \to \kappa_{0,c}} g_{\kappa_0,l}^{(2)}.
\eeq
This criterion does hold if the grand canonical ensemble
is dominated by large $N$ systems.
If the above criterion is satisfied for some value of $l$,
it does hold for larger values of $l$.
In the case of random walks, it already holds for the conventional
grand canonical ensemble of $l=0$ and we do not need to introduce
the generalized ensembles of $l >0$.

\par
From eqs.(\ref{RW:partition0}), (\ref{RW:correlation0}), the generalized
partition functions and the correlation functions are given by
\beqa
&& Z_{\kappa_0,l}= \left( \kappa_0 {\partial \over \partial \kappa_0}
\right)^l
                  Z_{\kappa_0},  \n
&& \hat{G}^{(2)}_{\kappa_0,l}(p) =
       \left( \kappa_0 {\partial \over \partial \kappa_0} \right)^l
       \hat{G}^{(2)}_{\kappa_0}.
\eeqa
We are interested in the behaviors
near the critical point $\delta \kappa \sim 0$ ($N \rightarrow \infty$)
and in the scaling region where
$1/(\kappa_c-\kappa_0) > 1/(\kappa_c - h(p) \kappa_0) \gg 1/\kappa_c$.
In this region, they become
\beqa
 Z_{\kappa_0,l} &\sim&
{V \kappa_c^{l+2} l! \over (\kappa_c-\kappa_0)^{l+1}}, \n
 \hat{G}^{(2)}_{\kappa_0,l}(p) &\sim&
{2 \kappa_c^{l+4} (l+1)! \over
(\kappa_c - h(p) \kappa_0)
(\kappa_c-\kappa_0)^{l+2}}
\left(
1 + {l \over l+1} {\kappa_c-\kappa_0 \over \kappa_c - h(p) \kappa_0} +
\cdots \right)
\n
&\sim&   {2 \kappa_c^{l+4} (l+1)! \over
(\kappa_c-\kappa_0)^{l+2}}
{1 \over  \kappa_0 H(p)}
\left(1 - \frac{\delta \kappa}{(l+1) \kappa_0}\frac{1}{H(p)}
+\cdots \right).
\eeqa
Since $\langle N \rangle =
Z_{\kappa_0,l+1}/ Z_{\kappa_0,l}=
\kappa_c (l+1) / \delta \kappa$ for the
generalized grand canonical ensembles of $l$,
the normalized correlation functions become
\beq
\hat{g}^{(2)}_{\kappa_0,l}(p)
\sim {2 \langle N \rangle \over V} {1 \over H(p)}
\left(1 -\frac{1}{\langle N \rangle H(p)} +\cdots \right)
\eeq
which agrees with the
previous result (\ref{RW:final}) of $l=0$.
\section{Branched polymer dynamics}
\setcounter{equation}{0}

\subsection{Canonical Ensemble}

Branched polymers are a statistical system of
$N$ points connected by $N-1$ bonds whose lengths are of order
$a_{0}$.
The canonical  partition function
is defined as
\beq
Z_N=\frac{1}{N!}\sum_{G:{\rm tree\;\; graph}} \int \prod_{i=1}^{N} d^d y^i
\prod_{(ij):{\rm bonf\;\;of\;\;}G} f(y^i-y^j),
\label{eq:cpf}
\eeq
where $f(x)$ is a function assigned to each bond in each graph, and
it damps sufficiently fast at long distances compared to the typical
length scale $a_{0}$.
The presence of the factor $1/N!$ is  due to the fact that the $N$ points
are regarded
identical.
\par
We can calculate a partition function for each $N$,
by counting the number of all possible tree graphs:
\beqa
Z_1&=& V, \n
Z_2&=& \frac{1}{2!}\hat{f}(0)V, \n
Z_3&=& \frac{1}{3!}3\hat{f}(0)^2 V, \n
Z_4&=& \frac{1}{4!}16\hat{f}(0)^3 V, \n
   &\vdots&\n
Z_N&=& \frac{1}{N!}N^{N-2}\hat{f}(0)^{N-1} V ,
\label{eq:Zn}
\eeqa
where $\hat{f}(p)$ is a Fourier transform of $f(x)$;
\beq
\hat{f}(p) = \int d^d x\; e^{ipx} f(x),
\eeq
and $V$ is the total volume of the system.
A derivation of the  general form (\ref{eq:Zn}) is given
in the appendix A.
\par
We define an (unnormalized)
$m$-point correlation function of density operators as
\beqa
&&G^{(m)}_N(x^1,\cdots,x^m) \n
&=& Z_N <\rho(x^1)\cdots\rho(x^m)>_N \n
&=&\frac{1}{N!}\sum_{G:{\rm tree\;\;graph}} \int \prod_{i=1}^{N} d^d y^i
\prod_{(ij):{\rm bond\;\;of\;\;}G} f(y^i-y^j)\;\;\rho(x^1)\cdots\rho(x^m),
\label{eq:ccf}
\eeqa
where the density operator is defined by
\beq
\rho(x) = \sum_{i=1}^N \delta^{(d)}(x-y^i).
\eeq
Due to the translational invariance,
one point function is proportional to the partition function;
\beq
G_N^{(1)} = {N \over V} Z_N.
\label{eq:g1nzn}
\eeq
The one-point function is nothing but the partition function
with one marked point.
\subsection{Grand canonical ensemble}
We then define  partition functions and  $m$-point correlation functions
in the generalized grand canonical ensembles as in section 2:
\beqa
Z_{\kappa_0,l}&=& \sum_{N=1}^{\infty}N^l \; \kappa_0^N \; Z_N,
\label{eq:gcpf}\\
G^{(m)}_{\kappa_0,l}(x^1,\cdots,x^m)
&=& \sum_{N=1}^{\infty}N^l \; \kappa_0^N \; G^{(m)}_N(x^1,\cdots,x^m).
\label{eq:gccf}
\eeqa
$\kappa_0$ is  the fugacity.

The criterion for a `good' grand canonical ensemble is such that
we can take the correct thermodynamic limit in the following sense.
The correlation functions in the grand canonical ensemble at the
critical value of fugacity should reproduce
those in the canonical ensemble for large $N$:
\beq
\lim_{N \to \infty } g_{N}^{(m)}
= \lim_{\kappa_0 \to \kappa_{0,c}} g_{\kappa_0,l}^{(m)},
\eeq
where we have defined  normalized correlation functions as
\beq
g_{N}^{(m)} = \frac{G^{(m)}_N}{Z_N}, \ \
g_{\kappa_0,l}^{(m)}= \frac{G^{(m)}_{\kappa_0,l}}{Z_{\kappa_0,l}}.
\eeq
In the case of random walks, we have confirmed that it does hold
for any nonnegative value of $l$ but we need to check it
in the case of branched polymers.
To satisfy this criterion,
the grand canonical ensembles with
$N$-dependent weights should be  dominated by large $N$ systems.
This is not assured only by taking the fugacity near the critical value.
This is because,
if $G_N^{(m)}$ behaves as $(\kappa_{0,c})^{-N} N^{\alpha}$ for large $N$
and $l+\alpha <-1$,
the summation over $N$
is dominated by small $N$ system, not by the large
$N \sim \kappa_c/\Delta \kappa$ even near the critical point.
On the other hand, if we take a sufficiently large $l$, we can
obtain the correct large $N$ correlation functions in the
grand canonical ensembles, which are, of course,  independent of $l$.
\par
We illustrate the above mentioned criterion by taking the partition function
as an example.
Since the canonical ensemble partition function (\ref{eq:Zn})
behaves at large N as
\beq
Z_N \sim \frac{N^{-5/2}}{\sqrt{2\pi} e^{-N}} \hat{f}(0)^{N-1} V,
\label{eq:znasym}
\eeq
the grand canonical ensemble is approximated by
\beqa
Z_{\kappa_0,l}&\sim& \frac{V}{\sqrt{2\pi} \hat{f}(0)}
       \sum_{N=1}^{\infty} N^{l-5/2} \;\; (\kappa/\kappa_c)^N \n
&\sim& \frac{V}{\sqrt{2\pi} \hat{f}(0)}
\int_0^{\infty} dN \;\; N^{l-5/2} \;\; e^{-N \Delta \kappa/\kappa_c},
\label{eq:integrand}
\eeqa
where
\beqa
\kappa&=&\hat{f}(0)\kappa_0, \n
\kappa_c &=& e^{-1},
\label{eq:kc1}\n
\Delta\kappa&=&\kappa_c - \kappa.
\eeqa
If we take $l$ sufficiently large,
the integrand in \eq{eq:integrand} has a peak at
$N \sim \kappa_c/\Delta \kappa$ and
 we can make $N$ large by letting $\kappa$ approach $\kappa_c$.
On the other hand, if $l$ is not sufficiently large, a non-universal
small $N$ behavior dominates the summation and we cannot obtain
the  correct answer of the large $N$ limit by a grand canonical ensemble.
\subsection{Schwinger-Dyson eq.}
In this subsection, we recapitulate the arguments that the
correlation
functions for branched polymers in the conventional grand canonical ensemble
are given by massless $\phi ^3$ theory.
Let's consider the following correlation functions
$G^{(m)}_{\kappa_0}= G^{(m)}_{\kappa_0, l=0}$,
which are suitable for 
Schwinger-Dyson analysis:
\beq
G^{(m)}_{\kappa_0} (x^1,\cdots,x^m) =
\sum_{N=1}^{\infty} \kappa_0^N G_N^{(m)}(x^1,\cdots,x^m)
\label{gengrand}
\eeq
We write a Fourier transform of $G^{(m)}_{\kappa_0}(x^1,\cdots,x^m)$
as $\hat{G}^{(m)}_{\kappa_0} (p^1,\cdots,p^{m-1})$:
\beqa
&&(2\pi)^d \delta^{(d)}(p^1+\cdots+p^m)
\hat{G}^{(m)}_{\kappa_0} (p^1,\cdots,p^{m-1}) \n
 && =
\int d^d x^1 \cdots d^d x^m \;\;e^{i p^1 x^1} \cdots e^{i p^m x^m}
\;\;G^{(m)}_{\kappa_0} (x^1,\cdots,x^m).
\eeqa

\begin{figure}[h]
\begin{center}
\leavevmode
\epsfxsize=6cm
\epsfbox{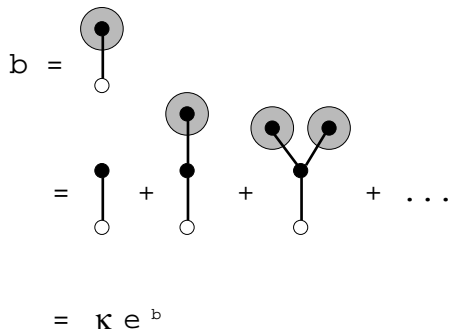}
\caption{Schwinger-Dyson equation for one point function.
A grey blob and a black point mean $\hat{G}^{(1)}_{\kappa_0}$ and
$\kappa_0$, respectively.
}
\label{fig:sd}
\end{center}
\end{figure}

\begin{figure}[h]
\begin{center}
\leavevmode
\epsfxsize=5cm
\epsfbox{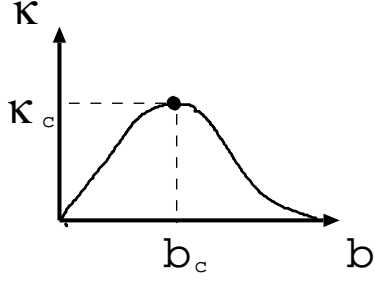}
\caption{Schwinger-Dyson equation, $\kappa = b e^{-b}$.
At the critical point, $b_c=1, \kappa_c=e^{-1}$.
}
\label{fig:bk}
\end{center}
\end{figure}

Schwinger Dyson equation for 1-point function $\hat{G}^{(1)}_{\kappa_0}$
becomes
\beq
b  = \kappa e^b,
\label{eq:sdb}
\eeq
where
\beq
b \equiv \hat{f}(0) \hat{G}^{(1)}_{\kappa_0},
\label{eq:defb}
\eeq
as can be seen from figure \ref{fig:sd}.
Figure \ref{fig:bk} illustrates \eq{eq:sdb}.

At the  critical point,
\beqa
b_c &=& 1, \n
\kappa_c &=& e^{-1},
\label{eq:kc2}
\eeqa
$\del b/\del k$ diverges.
Near this critical point, $N$ becomes large;
\beq
\Delta b \sim \sqrt{2e} \sqrt{\Delta \kappa} \sim 1/\sqrt{N},
\label{eq:delbsqrtn}
\eeq
where
\beqa
\Delta b =b_c-b, \n
\Delta \kappa =\kappa_c-\kappa.
\eeqa
The one-point function (which is equal to the
 partition function of $l=1$)
now behaves as follows.
\beqa
Z_{\kappa_0, l=1} &=&
\sum_N N \kappa_0^N Z_N = V \sum_N  \kappa_0^N G_N^{(1)}
= V \hat{G}_{\kappa_0}^{(1)} \n
&=& {b V \over \hat{f}(0)} \sim
\frac{V}{\hat{f}(0)}(1-\sqrt{2e}\sqrt{\Delta \kappa}).
\label{partition-function}
\eeqa
\par
\begin{figure}[h]
\begin{center}
\leavevmode
\epsfxsize=7cm
\epsfbox{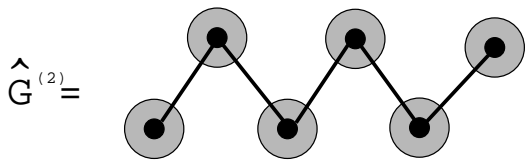}
\caption{2-point function $\hat{G}^{(2)}_{\kappa_0}$ is made out of
1-point function $\hat{G}^{(1)}_{\kappa_0}$, which is written by a gray
blob.
}
\label{fig:f2}
\end{center}
\end{figure}

Next, we consider the 2-point function $\hat{G}^{(2)}_{\kappa_0}(p)$.
When we pick up any two points on a tree graph,
we can fix the path connecting these two points.
Thus, as can be seen from figure \ref{fig:f2}, 2-point function is
calculated to be
\beqa
\hat{G}^{(2)}_{\kappa_0} (p)
&=& \sum_{s=0}^{\infty} \hat{f}(p)^s(\hat{G}^{(1)}_{\kappa_0})^{s+1} \n
&=&\frac{b }{\hat{f}(0)(1-b h(p))},
\label{2point}
\eeqa
where
\beqa
h(p)&\equiv& \hat{f}(p)/\hat{f}(0) \equiv 1 - H(p) \n
    &=&1-c a_0^2 p^2+\cdots.
\eeqa
Here $c$ is a positive constant of order one.
Recall that $f(x)$ damps rapidly out of the region $0<x<a_0$.

Near the critical point, $b \sim b_c=1$, the 2-point correlation
function behaves as
\beqa
\hat{G}^{(2)}_{\kappa_0} (p)
&=& \frac{b}{\hat{f}(0)} \frac{1}{H(p) + \Delta b h(p)}\n
             &\sim& \frac{1}{H(p) + N^{-1/2}}.
\eeqa
Here we used \eq{eq:delbsqrtn}.
Thus, the correlation length is $\xi = a_0 N^{1/4}$, which shows
the first definition of the Hausdorff dimension defined in (\ref{Hdim1})
to be  $d_H^{(1)}=4$.
Let us consider the following region:
\beq
a_0 \ll x <  \xi= a_0 N^{1/4}
\eeq
or in momentum space
\beq
{1 \over \xi} = N^{-1/4}/a_0 < p \ll 1/ a_0.
\label{eq:region}
\eeq
$a_0 $ gives an ultraviolet cut-off whereas
$\xi$ gives an infrared cut-off length over which correlation functions
damp rapidly.
In the  above scaling region of (\ref{eq:region}), the correlation function
behaves as an ordinary massless field
$\hat{G}^{(2)}_{\kappa_0}(p) \sim 1/p^2$ and  gives the second definition of
the Hausdorff dimension of (\ref{Hdim2}), $d_H^{(2)}=2$.
This is different from the Hausdorff dimension $d_H^{(1)}$ determined
>from the relation of the system size $N$ and the extent of the system.
In the next section, we show that the above behavior of
the correlation function in the scaling region is not correct
and hence gives the incorrect Hausdorff dimension $d_H^{(2)}$.
\par
Finally, we consider $m>2$ point correlation functions.
As in the case of the two-point function,
when $m$ points are fixed on each tree
graph, we can uniquely fix the path connecting these points.
Therefore, an $m$-point function $\hat{G}^{(m)}_{\kappa_0}$
is represented as a summation over
all tree diagrams with $m$ fixed points in which
$\hat{G}^{(2)}_{\kappa_0}(p)$ appear as propagators.
For example,
\beq
\hat{G}^{(3)}_{\kappa_0} (p,q)
= (\hat{G}^{(1)}_{\kappa_0})^{-2}
\hat{G}_{\kappa_0}^{(2)}(p)\hat{G}_{\kappa_0}^{(2)}(q)
  \hat{G}_{\kappa_0}^{(2)}(p+q).
\label{eq:santen}
\eeq
In general, we obtain the following result for $m$-point correlation
functions;
\beq
\hat{G}^{(m)}_{\kappa_0}
\sim \mbox{correlation functions of massless $\phi^3$ theory
at tree level}.
\label{eq:nvemfnc}
\eeq
In the next subsection, we consider generalized grand canonical
ensembles with larger $l$ and we point out that the results
(\ref{eq:nvemfnc})
do not correspond to the correct thermodynamic limit.

\subsection{Correlation functions in thermodynamic limit}
Let us consider the generalized $m$-point correlation functions with
$l \ge 1$.
From the definition (\ref{eq:gccf}), they can be obtained
by applying $l$-th derivative to the $l=0$ case:
\beqa
\hat{G}^{(m)}_{\kappa_0,l}(p^1,\cdots,p^{m-1})
&=&(\kappa_0 \frac{\del}{\del \kappa_0})^l
\hat{G}^{(m)}_{\kappa_0,l=0}(p^1,\cdots,p^{m-1})\n
&=&(\frac{b}{1-b} \frac{\del}{\del b})^l
\hat{G}^{(m)}_{\kappa_0,l=0}(p^1,\cdots,p^{m-1})
\eeqa

Partition functions  are obtained from (\ref{partition-function}) as
\beq
Z_{\kappa_0, l} = (\frac{b}{1-b} \frac{\del}{\del b})^{l-1}
{b V \over \hat{f}(0)}.
\eeq
Near the critical point ($b_c =1$), they become
\beq
Z_{\kappa_0, l \ge 2} \sim {V \over \hat{f}(0) }
{(2l-5)!! \over (1-b)^{2l-3}}.
\eeq
\par
2-point function with $l=1$ is given by
\beqa
\hat{G}^{(2)}_{\kappa_0,l=1}(p)
&=&(\frac{b}{1-b} \frac{\del}{\del b})
(\frac{b}{\hat{f}(0)(1-b h(p))}) \n
&\sim& \frac{1}{\hat{f}(0)} \frac{1}{(1-b)(1-bh(p))^2}\n
&\sim& \frac{1}{p^4}
\eeqa
It is because in the scaling region (\ref{eq:region}) and near the
critical point,
the following inequality holds:
\beq
\frac{1}{b} \sim 1
\ll \frac{1}{1-bh(p)} \sim \frac{1}{c (a_0 p)^2 + N^{-1/2}}
\ll\frac{1}{1-b} \sim \sqrt{N}.
\label{eq:ineq}
\eeq
This behavior is different from that of
$\hat{G}^{(2)}_{\kappa_0,l=0} \sim 1/p^2$.
Similarly for $l \ge 2$,  the behavior of 2-point function becomes
\beq
\hat{G}^{(2)}_{\kappa_0,l \ge 2}(p)
\sim
\frac{1}{\hat{f}(0)} \frac{(2l-3)!!}{(1-b)^{2l-1}\; (1-bh(p))^2}
[1+2 \frac{1-b}{1-bh(p)}+6\frac{l-2}{2l-3}(\frac{1-b}{1-bh(p)})^2
+\cdots].
\eeq
Due to the inequality (\ref{eq:ineq}),
the derivative $\frac{\del}{\del b}$
acts dominantly on
$\frac{1}{1-b}$, not on $\frac{1}{1-bh(p)}$.
The normalized 2-point functions now become
\beqa
\hat{g}_{\kappa_0, l \ge 2}^{(2)}(p)
&\sim& {1 \over V} \frac{2l-3}{(1-b)^2} \frac{1}{(1-h(p))^2}
(1-3\frac{(1-b)^2}{2l-3} \frac{1}{(1-h(p))^2}+\cdots) \n
&\sim& \frac{\langle N \rangle}{V} \frac{1}{H(p)^2}
(1-3\frac{1}{\langle N \rangle}\frac{1}{H(p)^2}+ O(({1 \over 
\langle N \rangle H(p)^2})^2)).
\label{eq:bp2pfgc}
\eeqa
Here we have used
$\langle N \rangle =
Z_{\kappa_0,l+1}/ Z_{\kappa_0,l}=
(2l-3) /(1-b)^2$ for the
generalized grand canonical ensembles of $l$.
Their $p$-dependences  are all the same except $l=0$ case.
The $l=0$ case, which can be obtained directly from the  Schwinger-Dyson
equation, does not reproduce the correct thermodynamic result.
Instead, we should consider a `good'
grand canonical correlation function
with $l \ge 1$, otherwise a non-universal small $N$ behavior
affects the summation and we cannot obtain the universal result.
In the appendix B, we estimate the large $N$ asymptotic behavior
of the two-point function by the saddle point method.
Such an explicit analytical result is completely
consistent with  the analysis here.
\par
The behavior of $\hat{g}^{(2)}_{\kappa_0,l \ge 1}(p) \sim 1/p^4$
gives (the second definition of ) the Hausdorff dimension (\ref{Hdim2})
$d_H^{(2)}=4$, which is now consistent with (the first definition
of) the Hausdorff dimension  discussed in the previous subsection.
An argument expected from the figure \ref{fig:f2} is that
the effect of branching could be absorbed by renormalizing
the mass. If so, the propagator would behave as that of random walks
with a renormalized mass and we might obtain the identical result with the
$l=0$
case.
We discuss in the next section why this argument is not correct.
\par
Similarly, 3-point functions become
\beqa
\hat{g}^{(3)}_{\kappa_0,l=0}(p,q) &\sim& g(p)g(q)g(p+q), \n
\hat{g}^{(3)}_{\kappa_0,l\ge 1}(p,q) &\sim&
g(p)'g(q)g(p+q)+g(p)g(q)'g(p+q)+g(p)g(q)g(p+q)',
\eeqa
where
\beqa
g(p)&=& \frac{1}{1-bh(p)} \sim \frac{1}{p^2}, \n
g'(p) &=& \frac{1}{(1-bh(p))^2} \sim \frac{1}{p^4}.
\eeqa
The behaviors do not change above $l=1$.
Only one propagator in a graph is replaced by $g'(p)$.
This is because
the derivative $\frac{\del}{\del b}$ dominantly act
on the factor $ 1/(1-b)$ rather than on
$1/(1-b h(p))$, as in the case of 2-point functions.
Here again, we might be tempted to argue
that the only effect of branching
is mass renormalization and to give rise to a three point vertex.
If so, we might conclude that
the  3-point function  behaves as the result of $l=0$ case.
As we discuss in detail in the next section, this
argument is not correct
in the thermodynamic limit and we should only retain $l \ge 1$ cases.
Hence correlation functions for branched polymers are expressed
in terms of $\phi^3$ theory at tree level with a single mass insertion.
\par
For $m$-point functions ($m>3$), we can obtain the same result.
\beqa
\hat{g}^{(m)}_{\kappa_0,l=0} &\sim& \mbox{correlation functions for
$\phi^3$ theory at tree level} \label{eq:wrong} \n
\hat{g}^{(m)}_{\kappa_0,l \ge 1} &\sim&
\mbox{correlation functions for $\phi^3$ theory at tree level
with a mass insertion}. \n
\label{eq:correct}
\eeqa
The universal correlation functions with $l \ge 1$ represent
the correct correlation functions in the thermodynamic limit.

As a consistency check,
the following relation between
an ($m$+1)-point function and an $m$-point function must hold:
\beq
\hat{g}^{(m+1)}_{\kappa_0,l \ge 1}(p^1,\cdots ,p^{m-1} ,p^m =0)
=N\hat{g}^{(m)}_{\kappa_0,l \ge 1} (p^1,\cdots ,p^{m-1}).
\label{eq:mmpo}
\eeq
It actually holds because in the L.H.S. of \eq{eq:mmpo},
the special class of
diagrams dominate in which the $m$-th end point is attached to
a propagator $g'(p)$.
It is due to the inequality $g'(p=0)g(p) \gg g'(p)g(p=0)$.
Therefore, it is equal to the R.H.S. of \eq{eq:mmpo}.
\section{Physical interpretation by a single mother universe}
\setcounter{equation}{0}
In this section, we give a physical interpretation why
the propagator behaves as $1/p^4$ instead of
$1/p^2$.
As we can see from the figure \ref{fig:f2},
the effect of branching seems to be absorbed by renormalizing
the mass
and we might conclude that the two-point function behaves
as that of random walks.
If this is the case,
the propagator should be given by an ordinary massive scalar
particle with a renormalized mass $(a_0 N^{1/4})^{-1}$,
instead of $(a_0 N^{1/2})^{-1}$.
Similarly, higher point correlation functions should be
given by tree graphs of $\phi^3$ field theory.
As we saw in the previous section, these are not the correct
behaviors of correlations.
In this section, we give a physical interpretation why
the propagator behaves as $1/p^4$ instead of a conventional behavior
$1/p^2$ and why the higher point correlation functions behave
as $\phi^3$ theory with a single mass insertion.
\par
Let us consider the two-point function as an example.
Two-point function  $\hat{G}^{(2)}_{\kappa=0}(p)$
was defined by a sum of two-point
functions in the canonical ensemble (eq.(\ref{gengrand})):
\beq
\hat{G}^{(2)}_{\kappa_0} (p)=
\sum_{N=1}^{\infty} \kappa_0^N \hat{G}_N^{(2)}(p).
\eeq
On the other hand, from the equation (\ref{2point}), it is written
as a sum of all contributions over $s$ where $s$ is the length between
the two points in concern.
Each term $\hat{G}^{(1)}_{\kappa_0}$ represents a gray blob in
figure \ref{fig:f2}.
When we fix the total number of the system $N$, the $N$ points
are distributed among $(s+1)$ blobs.
We will show here that the most dominant contributions to the
correlation functions are those that most of $N$ points are
concentrated on only a single blob. We call this blob the mother
universe. In the branched polymer, there is only
one mother universe and the other universes (blobs) contain much
fewer points than the mother universe.
\par
To show this, we first note that the 1-point function (blob)
is expanded as
\beq
\hat{G}^{(1)}_{\kappa_0} = \sum_{n=1}^{\infty} \kappa_0^n u_n
\eeq
where $u_n$ is given at large $n$ as
\beq
u_n = G_n^{(1)}
\sim {n^{-3/2} \hat{f}(0)^{n-1} e^n \over \sqrt{2\pi} }.
\label{uofn}
\eeq
By using this expansion and eq.(\ref{2point}), we obtain the
two-point function for fixed $N$ as
\beq
\hat{G}_N^{(2)}(p)=
\sum_{s=0}^{\infty} \hat{f}(p)^s
\left(
\sum_{n_0=1}^{\infty} \cdot  \cdot \cdot \sum_{n_s=1}^{\infty}
u_{n_0} \cdot  \cdot \cdot u_{n_s}
\delta_{N, n_0+ \cdot  \cdot \cdot n_s}
\right).
\eeq
Each contribution in the bracket comes from a graph in which
the first blob contains $n_0$ points, the second $n_1$ points,
and so on.
Using (\ref{uofn}), the term in the bracket becomes

\beq
\left( {1 \over \sqrt{2 \pi}} \right)^{s+1} \hat{f}(0)^{N-s-1} e^{N}
\sum_{n_0=1}^{\infty} \cdot  \cdot \cdot \sum_{n_s=1}^{\infty}
n_0^{-3/2} n_1^{-3/2} \cdot  \cdot \cdot n_s^{-3/2}
\delta_{N, n_0+ \cdot  \cdot \cdot n_s}.
\eeq
In the case of $s=1$, the summation
\beq
\sum_{n=1}^{N-1} n^{-3/2} (N-n)^{-3/2}.
\eeq
is dominated by terms of $n \sim 0$ and $n \sim N$.
For the general case with the exponent $\alpha$,
\beq
\sum_{n=1}^{N-1} n^{\alpha} (N-n)^{\alpha},
\eeq
the sum is dominated at the boundaries for $\alpha < -1$
and asymmetry arises between two blobs.
On the contrary, for $\alpha > -1$,
$N$ points are distributed equally and neither  blob
is special.
This argument can be generalized to $s>1$.
To conclude, most of $N$ points belong to a single
blob along the propagator.
We then have to divide the propagator with length $s$ at the mother
universe. Since the other blobs contain only a finite number of
points, the effect of branching other than dividing the propagator
into two pieces is simply to renormalize the mass
of propagator.
Hence, the propagator behaves as a product of two ordinary ones with
a renormalized mass;
\beq\
\hat{G}^{(2)}_N(p) \sim \left( {1 \over p^2 + m^2} \right)^2.
\eeq
\par
We can also apply a similar argument for the higher-point functions.
If the effect of branching is only to renormalize the mass term,
the higher-point correlation functions will be represented by
diagrams of $\phi^3$ theory with propagators
$1/(p^2 + m^2)$, which is not the correct thermodynamic behavior
of the correlation functions as we saw in the previous section.
Similar to the case of 2-point functions above,
we can argue that
there is only one mother universe in which most of the $N$
points reside.
Since only a single blob contains an infinitely many points,
we have to divide one of the propagators at the mother universe
and this propagator behaves as $1/(p^2+m^2)^2$.
This mother universe corresponds to the mass insertion.
The other blobs contain a finite number of points and
the effect of branching can be absorbed into mass renormalization.
This is the physical reason why the higher point correlation functions
are represented by tree diagrams of $\phi^3$  field theory
with a single mass insertion.
\par
The above statement that there is only one mother universe in the
branched polymer
is explained differently as follows.
Let us consider again the original branched polymer systems
with $N$ points and $N-1$ bonds.
By counting the number of ways in which we can divide
a branched polymer into two parts by cutting a bond,
we obtain a relation
\beq
(N-1) Z_N/V  = \hat{f}(0) \sum_{N'=1}^{N-1} \;\; [N' Z_{N'}/V] \;\;
[(N-N')Z_{(N-N')}/V].
\label{eq:cuteq}
\eeq
The factor $(N-1)$ in the L.H.S.
is interpreted as the number of bonds we can cut
to divide the whole into two parts.
The factors $N'$ and $(N-N')$ in the R.H.S.
are interpreted as the number of
points to which the bond connecting the two parts is attached.
Since $Z_N$ behaves as in eq.(\ref{eq:znasym}) at large $N$,
\beq
N Z_N/V  \gg \hat{f}(0) \sum_{N'=N\epsilon}^{N(1-\epsilon )} \;\; [N'
Z_{N'}/V] \;\;
[(N-N')Z_{(N-N')}/V],
\label{eq:cutineq}
\eeq
which means the summation in the R.H.S. of eq.(\ref{eq:cuteq})
is dominated by terms of
$N'\sim 0$ or $N'\sim N$.
This formula can be interpreted as follows.
If we divide any  graph into two parts by cutting some bond,
we find only finite points in one of them and
most of them belong to the other.
The mother universe belongs to the larger part.
Consider the 2-point function as an example.
By dividing  the graph of figure \ref{fig:f2}
into two parts at some bond, we find
that most points are dominantly distributed on
only one of them.
\par
We can  apply the same procedure to the dominant part
with infinite points.
After repeating it several times to divide  the total graph into
several pieces, we find that only  a single part consists of
infinitely many points and the others consist of finite points.
We can find out which blob in figure \ref{fig:f2} is
the mother universe
when, after the several repetitions,
the dominant part with infinite points is detached from
the path in figure \ref{fig:f2}.
We can also apply the same argument to the higher-point functions.
Here we point out the difference of the argument given here and
that for random surfaces\cite{kawai}.
Although a similar inequality appears
for random surfaces, the factor $N$ is absent on the
L.H.S. of the inequality (\ref{eq:cutineq}) in that case.

\section{Conclusion and discussion}
\setcounter{equation}{0}

In this paper we have shown that
the correlation functions for branched polymers are given by
those for $\phi^3$ theory at tree level {\bf with a single  mass insertion}
if we correctly take the thermodynamic limit.
It is not given by those for $\phi^3$ theory at tree level themselves.
We have interpreted the single mass insertion as the presence
of a single mother universe in branched polymers.

We have reviewed random walks in order to clarify
the relation between the canonical
and grand canonical ensemble. We have introduced generalized
 grand canonical
ensembles
which assign different weights for different system sizes.
We have emphasized that a `good' grand canonical ensemble is such that the
ensemble
average should be dominated by the systems of large size.
In branched polymers, this criterion is not satisfied in the conventional
grand
canonical ensemble. Nevertheless we can consider `good' grand canonical
ensembles in
branched polymers. Our conclusion follows 
as the universal prediction of `good'
grand canonical ensembles. It represents the correct scaling behavior of the
correlation functions in canonical ensembles of large system size $N$.

\par
\begin{center} \begin{large}
Acknowledgments
\end{large} \end{center}
This work is supported in part by the Grant-in-Aid for Scientific
Research from the Ministry of Education, Science and Culture of Japan
and by the National Science Foundation under Grant No. PHYS94-07194.
\section*{Appendix A}
\setcounter{equation}{0}
\renewcommand{\thesection}{A}
In this appendix, we derive the canonical ensemble partition function
(\ref{eq:Zn}) from the Schwinger-Dyson equation (\ref{eq:sdb}).
Let us solve $b$ as a form of expansion in $\kappa$.
Each coefficient is calculated to be
\beqa
\frac{1}{2\pi i} \oint_{\kappa=0} d\kappa \frac{b}{\kappa^{N+1}}
&=& \frac{1}{2\pi i} \oint_{b=0} db (1-b) e^{-b}
\frac{b}{(b e^{-b})^{N+1}}\\
&=& \frac{N^{N-1}}{N!}.
\eeqa
Hence,
\beq
b=\sum_{N=1}^{\infty} \frac{N^{N-1}}{N!} \kappa^{N}.
\eeq
From eqs.
(\ref{eq:g1nzn}), (\ref{gengrand}) and (\ref{eq:defb}), $b$ is expanded as
\beq
b=\hat{f}(0)\; \sum_{N=1}^{\infty} \frac{N}{V}\;
Z_N \; \kappa_0^N.
\eeq
Comparing these two expansions, we get the result of \eq{eq:Zn}.
\section*{Appendix B}
\setcounter{equation}{0}
\renewcommand{\thesection}{B}

In this appendix, we derive the two-point correlation function
in the canonical ensemble for fixed but large $N$.
Similar to  the calculation in appendix A,
we can obtain the two-point function
for a fixed $N$ from eq.(\ref{2point});
\beqa
\left( {1 \over \hat{f}(0) } \right)^N G_N^{(2)}(p) &=&
\frac{1}{2\pi i} \oint_{\kappa=0} d\kappa
\frac{\hat{G}^{(2)}(p)}{\kappa^{N+1}} \n
&=&  \frac{1}{2\pi i} \oint _{b=0} db
{1\over \hat{f}(0) }{1-b \over 1-b h(p)} b^{-N} e^{b N}.
\eeqa
This integration can be estimated for large $N$ by the saddle point
approximation.
Since $b^{-N} e^{b N} = e^{N(b- \log  b)}$,
the saddle point is at $b=1$. The steepest descent direction
is along the imaginary direction.
We change the variable from $b$ to $t$ around the saddle point as
\beq
b - \log  b = 1 -{t^2 \over 2}.
\label{saddleparameter}
\eeq
The new parameter $t$ is written
in terms of $b$ by
\beq
t = i(1-b) (1+ {1-b \over 3} + {7 \over 36} (1-b)^2 + \cdot \cdot \cdot).
\eeq
Solving this, we obtain
\beq
{1 \over b} -1 = -i t (1- {2 \over 3} it - {13 \over 36} t^2
+ \cdot \cdot \cdot).
\eeq
Therefore in the saddle point approximation, the correlation function
becomes
\beqa
&& \left( {1 \over \hat{f}(0) } \right)^{N-1} \hat{G}_N^{(2)}(p) =
\frac{1}{2\pi i}
\int_{- \infty}^{\infty} dt {t \over 1/b -1 + H(p)}
e^{N(1-{t^2 \over 2})}
\n
 &=&  \frac{e^N}{2\pi}  \int_{- \infty}^{\infty} dt
{t^2 (1- 13 t^2/36  + \cdot \cdot \cdot)
\over
(H(p)-2 t^2/3 + \cdot \cdot \cdot)^2 +  t^2 (1- 13 t^2/36
+ \cdot \cdot \cdot )^2 }
e^{-N t^2 /2}
\n
&=& {N^{-3/2} e^N \over 2 \pi}
 \int_{- \infty}^{\infty} dt
{t^2 (1- 13 t^2/36N  + \cdot \cdot \cdot)
\over
(H(p)-2 t^2/3N + \cdot \cdot \cdot)^2 +
t^2/N (1- 13 t^2/36 N + \cdot \cdot \cdot )^2 }
e^{- t^2 /2}
\eeqa
where $H(p)= 1- h(p) \sim c(a_0 p)^2 $.
Hence the normalized correlation function becomes
\beqa
&& \hat{g}_N^{(2)}(p) = {\hat{G}_N^{(2)}(p) \over Z_N } \n
&& = {N \over V}
 \int_{- \infty}^{\infty} {dt \over \sqrt{2 \pi} }
{t^2 (1- 13 t^2/36N  + \cdot \cdot \cdot)
\over
(H(p)-2 t^2/3N + \cdot \cdot \cdot)^2 +
t^2/N (1- 13 t^2/36 N + \cdot \cdot \cdot )^2 }
e^{- t^2 /2} \\
&&= {N \over V}  \int_{- \infty}^{\infty} {dt \over \sqrt{2 \pi} }
{t^2 \over H(p)^2 -4t^2 H(p)/3N+ t^2 /N} e^{- t^2 /2}
(1 + O({1 \over N})).
\eeqa
The correlation function for branched polymers behaves very
differently from the case of simpler random walks.
In the scaling region $1/\sqrt{N} <  H(p) \ll 1$,
the 2-point function can be evaluated as
\beqa
\hat{g}_N^{(2)}(p)
&\sim&
{N \over V}  \int_{- \infty}^{\infty} {dt \over \sqrt{2 \pi} }
{t^2 \over H(p)^2 + t^2 /N} e^{- t^2 /2}
 \eeqa
This result for canonical ensemble is completely consistent with the
result of the grand canonical ensemble (\ref{eq:bp2pfgc}).

\end{document}